\documentclass[preprint]{aastex}
\input psfig.sty

\def\H7{\mbox {$h_{0.7}$}}
\def\lstarlya{\mbox {$L^*_{Ly\alpha}$}}
\def\IZw18{I~Zw~18}
\def\m82{M82}

\def\h{\mbox {\rm H}}

\def\deg{\mbox {$^{\circ}$}}
\def\msun{\mbox {${\rm ~M_\odot}$}}

\def\msunyr{\mbox {$~{\rm M_\odot}$~yr$^{-1}$}}
\def\angs{\mbox {~\AA}}
\def\lya{\mbox {Ly$\alpha$~}}
\def\Ha{\mbox {H$\alpha$~}}

\def\lstar{\mbox {$L^*$}}

\def\h0{\mbox {~H$_0$}}
\def\q0{\mbox {~q$_0$}}

\def\asec{\ifmmode {'' }\else $''~$\fi}  
\def\amin{\ifmmode {' }\else $'~$\fi}    
\def\arcsper{\ifmmode \rlap.{'' }\else $\rlap{.}'' $\fi} 
\def\arcmper{\ifmmode \rlap.{' }\else $\rlap{.}' $\fi} 
\def\sles{\lower2pt\hbox{$\buildrel {\scriptstyle <}
   \over {\scriptstyle\sim}$}} 
\def\sgreat{\lower2pt\hbox{$\buildrel {\scriptstyle >}
    \over {\scriptstyle\sim}$}} 
\def\kms{~km~s$^{-1}$~}
\def\ergsec{~ergs~s$^{-1}$~}

\def\flux{~ergs~s$^{-1}$~cm$^{-2}$}

\def\cm3{~cm$^{-3}$}
\def\mpc3{~Mpc$^{3}$}
\def\mpc-3{~Mpc$^{-3}$}

\def\fig{{Figure}}

\def\et{{\rm et\thinspace al.}\ }   

\def\apj{ApJ}
\def\apjs{ApJS}
\def\pasp{PASP}
\def\aj{AJ}
\def\mn{MNRAS}

\def\aa{A\&A}

\begin{document}
\title{The Space Density of Redshift 5.7 \lya\ Emitters: 
First Constraints from a Multislit Windows Search\altaffilmark{1}}

\author{Crystal L. Martin\altaffilmark{2,} \altaffilmark{3}}
\affil{University of California Santa Barbara}
\affil{Department of Physics}
\affil{Santa Barbara, CA 93106}
\email{cmartin@physics.ucsb.edu}

\author{Marcin Sawicki\altaffilmark{4}}
\affil{Dominion Astrophysical Observatory} 
\affil{Herzberg Institute of Astrophysics}
\affil{National Research Council}
\affil{5071~West Saanich Road}
\affil{Victoria, B.C., V9E 2E7,Canada}

\email{marcin.sawicki@nrc.gc.ca}

\altaffiltext{1}{Data presented herein were obtained at the W.M. Keck
Observatory, which is operated as a scientific partnership among the 
California Institute of Technology, the University of California and the
National Aeronautics and Space Administration.  The Observatory was made
possible by the generous financial support of the W.M. Keck Foundation.
}
\altaffiltext{2}{Packard Fellow}
\altaffiltext{3}{Alfred P. Sloan Research Fellow}
\altaffiltext{4}{Guest User, Canadian Astronomy Data Centre, which is operated by the Herzberg Institute of Astrophysics, National Research Council of Canada.}

\author{Revised Version of 2003 September 21}

\begin{abstract}
We present results from a blind, spectroscopic search for redshift 5.7
\lya\ emission-line galaxies at Keck I. Using a band-limiting filter
and custom slitmask, we cover the LRIS detector with low resolution 
spectra in the $8100 - 8250\angs$ atmospheric window which contains 
no bright night sky emission lines. We find nine objects with line fluxes 
greater than our flux limit of $6 \times 10^{-18}$\flux\ in our
$\sim 5.1$ square arcminute field.  We rule out a
\lya\ identification for six of these based on the absence of the continuum 
break, expected at rest-1215 \AA\ for high-z galaxies, and/or the
identification of additional emission-lines in our follow-up spectra.
We find that extremely metal-poor, foreground emission-line
galaxies are the most difficult type of interloper to recognize.
For the three remaining emission-line objects,  we identify a plausible 
counterpart for each object in a  deep V-band image of the field suggesting 
that none of them has a continuum break in the $i$ band. 
Our preliminary conclusion is that our field contains
no z=5.7 \lya\ emitters brighter than $0.6\lstarlya$, where $\lstarlya\
\equiv 3.26 \times 10^{42}$\ergsec.  Selecting a field with zero \lya\ 
emitters is marginally consistent with the no-evolution hypothesis -- i.e.
we expected to recover 2 to 3 \lya\ emitters assuming the  \lya\ luminosity 
function at redshift 5.7 is the same as it is at redshift 3.
Our null result rules out 
a brightening of \lstarlya\ by more than a factor of 1.7 from redshift 3
to redshift 5.7, or, over the same redshift interval, 
an increase of more than a factor of 2.2 in the number density of
Ly$\alpha$ emitters.
The paucity of z=5.7 \lya\ emitters raises the question
of whether the \lya -selected population plays a significant role in
maintaining the ionization of the intergalactic medium at $z = 5.7$.
We find that if the escape fraction of \lya\ radiation is  less than
$0.4 f_{LyC}$, where $f_{LyC}$ is the escape fraction of Lyman continuum
photons, then the star formation rate in the \lya\ emitting
population  is high enough in the no-evolution model (our upper limit)
to maintain the ionization of the IGM at z=5.7.

\end{abstract}

\subjectheadings{
galaxies: formation --- 
galaxies: evolution ---
galaxies: high-redshift ---
techniques: spectroscopic
}

\section{Introduction}

The Epoch of Reionization likely marks a distinct change in the galaxy
population in addition to a phase transition in the intergalactic
medium.  Sustained star formation is probably prevented by supernova
feedback in the first, low mass galaxies. Completion of
Reionization likely follows the formation of fairly massive
galaxies that are immune to such violent feedback.
The WMAP satellite has measured
a large optical depth to electron scattering after cosmological
recombination implying significant reionization at $z \sim 17 \pm 5$
(Kogut \et 2003; Spergel \et 2003).  Yet the most recent re-heating
must have occured more recently based on the thermal history of the
intergalactic medium derived from the \lya\ forest (Wyithe \& Loeb
2003).  The absence of neutral regions in the intergalactic medium at
$z \sles\ 6$ (Becker \et 2001; Djorgovski \et 2001) indicates
Reionization was completed by this time.  The population of galaxies
responsible for Reionization, however, has yet to be recognized.

Searches for $z \sim 6$ objects latch onto either the \lya\ emission line or 
the Lyman continuum break. Over cosmological distances, the HI opacity of the 
IGM blankets the continuum from 912\angs\ to the \lya\ line 
effectively shifting the continuum break up to 1216\AA\ (Madau 1995). 
For galaxies at redshift 5 to 7,  this break crosses the $r$ and $i$ bands, 
producing very red $r-i$ galaxy color (or, as redshift increases, $i-z$ color).
Selection by very red color, i.e. r-band (or i-band) dropouts, provides
many high-redshift candidates (e.g. Yan \et 2003) but is not a sufficient 
selection criterion.  Galactic L and T dwarf stars, as well as 
$z \sim 1$ galaxies with a 4000\angs\ continuum break,  have similar red 
colors in these bands.  Broad-band
selection followed by spectroscopic follow-up for \lya\ emission has provided
some of the best statistics on high-z galaxies to date (Lehnert \& Brewer
2003), but the spectroscopic follow-up required for confirmation 
is difficult. Indeed, to date, the highest redshift galaxies have been 
discovere fd in narrow-band imaging surveys for \lya\ emission (z=6.56, Hu 
\et 2002; z=6.541, 6.578, Kodaira \et 2003).  Selection by \lya\ emission,
however, is clearly not complete as only $\sim 25\%$ of 
starburst galaxies at $z \sim 3$ show strong \lya\ line emission
(e.g. Steidel \et 2000; Shapley \et 2003). 
Since line selection and continuum-break selection have different
shortcomings, progress will likely be made by using the two
techniques in parallel.

At $z \sgreat\ 5$, a method for detecting intrinsically faint galaxies 
-- i.e. typical galaxies rather than rare objects -- in large numbers 
is clearly needed to characterize their number density and clustering 
properties. Surveys for lensed objects along cluster caustics probe
extremely deep but search tiny volumes  (e.g. Ellis \et 2001). In contrast,
narrowband imaging covers large  areas on the sky but does not go as
deep as desired.
The survey described in this paper explores 
a technique that may allow us to probe further down the luminosity function
than narrowband imaging surveys while covering more area than lensing
surveys.  This multi-slit windows technique
minimizes the sky noise under an emission line by dispersing the
light but covers significant area on the sky by  stacking multiple longslits
on a single mask.  We describe the strategy and our pilot observations
in \S 2 of this paper.  In \S 3, we present our catalog of emission-line
objects and discuss the line identifications.  We constrain the number
density and evolution of high redshift \lya\ emitters in \S 4.  Section
5 summarizes our main results and the utility of the multi-slit windows
technique.
We adopt an $\Omega_{m} = 0.3$, $\Omega_{\Lambda} = 0.7$, $h = 0.7$
cosmological model throughout this paper.

\section{Emission-Line Search}

To search for \lya\ emission line objects at redshift 5.7, 
we have carried out a {\it multislit windows} search at Keck~I with LRIS 
(Oke \et 1995) over the  atmospheric window at 8200\angs. Dispersing the light
with a grating maximizes the line vs sky contrast.
A narrowband filter restricts the spectral bandpass to the 
airglow-free region illustrated in Figure~\ref{fig:skywindow}.
Because the spectra are only $\sim 100$ \AA\ wide, we can cut multiple 
long-slits on a single slit-mask, thereby greatly increasing the sky area 
surveyed.  Similar techniques have been described by Crampton and Lilly (1999) 
and Stockton (1999) in the atmopheric window at 9200\AA
Our aim was to systematically probe 
the 8200\angs\  window for  \lya\ emission at z=5.7 going
significantly deeper in intrinsic luminosity than the Crampton \& Lilly work
and narrowband-imaging surveys. Two key features of our strategy were
(1) near real-time spectroscopic follow-up of  emission-line candidates,
and (2) a quantitative analysis of our selection function
based on simulations of our detection efficiency.

\subsection{Experimental Technique}

The shape of the filter transmission profile is an important aspect of
the experimental design. 
 To optimally use the detector area, the filter transmission
curve should have a top-hat profile with width less than or equal to 
that of the atmospheric bandpass.
Figure~\ref{fig:skywindow} shows the filter transmission curve of
NB8185 which Esther Hu kindly loaned to us.  
The transmission profile is centered at  8197\angs\ and
 has a full width at half maximum intensity
(FWHM) of 106\angs\ and full width at 10\% transmissivity of 145\angs.
The filter wings transmit bright sky lines, so we spaced the
spectra 165\angs\ apart  to avoid contaminating the targetted bandpass
with sky lines from neighboring slits.  The increased slit spacing, over
that allowed by a perfect top-hat transmission profile, effectively 
reduces the survey area by a factor of 1.56.

We designed our mask for the Keck~I LRIS instrument and the NB8185 filter.
Our selection of the 150 ${\rm mm}^{-1}$ grating (4.7 \angs\
pix$^{-1}$) reflects a tradeoff between maximum sensitivity (somewhat
higher spectral resolution) and survey volume (more slits at lower
resolution). Rest-frame line widths of redshift 5.7 \lya\ emitters are 
260\kms\ (Bunker \et 2003) to 340\kms\ (Ajiki \et 2002), so the ideal 
spectral reslution for our experiment is 7 to 10 \angs. We chose a
slit width of 1\farcs5 which projects to 4.7 pixels and yields a
spectral resolution of about 22\angs\ for a source filling the
slit.  At this dispersion,  we fit thirty-three  slitlets onto the mask 
at 165\angs\ spacing. \fig~\ref{fig:mask} shows the mask layout.
Each of the longslits projects onto the full 7\farcm27 length of the 
Tektronix $2048 \times 2048$ CCD. The combined slitlets subtend 5.09 square 
arcminutes on the sky. Stability bars
prevent the thin aluminum mask from buckling; and small holes
allow accurate positioning of the mask using foreground stars.
The mask design does not cover the full LRIS field due to vignetting
in the spectrograph, and the spectra cover just over 50\% of the 
detector width in the dispersion direction.

The effective search area for faint objects, however, is significantly lower 
than the area subtended by the slits. Partial occultation by the mask 
renders objects with fluxes near the limiting flux of the survey too
faint to detect unless they land in the center of a slit.  These slit losses 
are derived quantitatively in \S 2.3. The depth of our search
along the sightline, set by the filter bandpass, also has a slight dependence 
on flux since we measure the product  $F T(\lambda)$ where $F$ is the line
flux and $T(\lambda)$ is the filter transmission at the wavelength of the line.
We include this effect in our modeling of the source counts.
The sky noise (and therefore our sensitivity) was found to be
fairly uniform across a bandpass equal to one filter FWHM. For a 1\farcs5 
(or 1\farcs0) effective slit width, our survey volume is 1100 (or 730) Mpc$^3$.

We choose a field centered at RA = 0:26:35.86, DEC = 17:11:35.15,
  J2000.0 This field covers the core of the rich cluster CL0024+16,
  redshift 0.397 (Czoske 2001).  The field was chosen because it could
  be observed at a low airmass from Hawaii, the reddening is low
  $E_{B-V} = 0.057$ (Schlegel, Finkbeiner, \& Davis 1998), and deep
  broadband imaging had been obtained previously for the galaxy cluster.  
  By selecting a cluster field, we also improve the chances of
  serendipitously finding a lensed object.  
  




\subsection{Observations}


We obtained the search spectra  2002 November 6 with the
Low-Resolution Imaging Spectrometer (Oke \et 1995) on Keck~I.
A total of 21,600 s of integration time was obtained under clear
skies. A series of exposures were made to allow us to track flexure, remove 
cosmic rays, and apply a 4\arcsec\ dither along the slits.  Mask
alignment was verified on alternating exposures by checking the flux
of the alignment stars.  The median seeing measured from
the spatial profiles of the alignment stars was 0\farcs8.


Calibration followed standard procedures using IRAF\footnote{The Image
  Reduction and Analysis Facility is written and supported by the IRAF
  programming group at the National Optical Astronomy Observatories
  (NOAO) in Tucson, Arizona. NOAO is operated by the Association of
  Universities for Research in Astronomy (AURA), Inc. under
  cooperative agreement with the National Science Foundation.}.  The
  mean bias level of each frame was measured for each amplifier and
  subtracted from the detector section read by that amplifier.  
  Pixel-to-pixel variations in gain were flattened by dividing each
data frame by an appropriately normalized exposure of a quartz lamp.
The images were aligned using integer shifts
to compensate for instrument flexure and dithering.
The 12 frames were averaged using cosmic ray rejection.
  The dispersion and spectral coverage were
  verified using Ar lines in an exposure of a calibration lamp.  At
  such low resolution, it was not possible to directly measure the
  spectral resolution as all lines are blends. 

We derived the flux calibration from observations of 
spectrophotometric standards made with a 1\farcs5 wide longslit.  
Despite the presence of an atmospheric absorption band in
the filter bandpass, the two stars we observed, Feige~34 and Hiltner~600 
(Massey \& Gronwall 1990), gave remarkably consistent
values  for the sensitivity (within a few percent).  The systematic 
errors introduced by the unknown location of a source within the
slit are much larger and are discussed later in this paper.
During the portion of the night with no moon, 
the measured sky brightness in our bandpass was
$3.7 \times 10^{-18}$\flux \angs $^{-1}$.

Cleaning the image with sky and continuum subtraction was necessary to
make automated searching possible.  Although the brightest atmospheric 
emission lines have been blocked out of our spectra, we subtracted local sky
in the combined spectral image by using a running sky window that
subtended a region from 1\farcs06 to 3\farcs20 in both directions
along the spatial axis.  The
resulting S/N in the sky subtracted frame reaches the theoretical
limit set by the shot noise from sky photons, so the amplitude of the random
noise fluctuations varies with wavelength as the square root of the
intensity of the sky spectrum.  We fit the continuum line by line for each 
panel of the sky-subtracted image and subtracted the fit.
Emission-lines can be picked out of this cleaned image
in a manner similar to object detection in direct images.

\subsection{Simulations}

We used simulations to gauge the sensitivity of our emission line
search.  We inserted artificial emission lines into our
stacked and cosmic-ray cleaned -- but yet not sky- or continuum-subtracted --
data frames. 
We used artificial emission lines with Gaussian spatial 
and spectral profiles. The three  combinations of  parameters
were: (FWHM$_{spatial}$, FWHM$_\lambda$) = (0.85\arcsec, 18.4\AA),
(1.07\arcsec, 23\AA), and (1.28\arcsec, 30\AA).    These choices
were motivated by the expected velocity dispersion -- 
rest-frame $220$\kms~FWHM (observed line width of 6\angs)
and our spectral resolution -- 22\angs\ for an object filling the slit.
Since more compact objects are easier to detect, these cases include
the limit of an unresolved source but also
explore the sensitivity to source profile.
  For each line shape, we ran simulations for a 
range of line brightnesses from 75 to 1000 digital units
(which corresponds to $2.91 \times 10^{-18} {\rm ~to~}
3.89 \times 10^{-17}$\flux\ for no slit loses).
We avoided placing artificial emission lines on bright
continuum objects, just as we avoided regions around such objects in
our automated search.


We automated the recovery of these artificial sources by tuning the parameters 
in the SExtractor software (Bertin \& Arnouts 1996). We choose 
regions with at least 4 connected pixels with S/N exceeding 0.5 times
the rms background fluctuations.    The 
noise level was computed in a ring around a candidate, so it sampled
both the sky and continuum noise.  This approach proved more robust
than the human eye which may be drawn to regions of high shot
noise where bright sky lines intersect continuum sources.
The sky aperture underestimates the noise in regions near the
edges of the bandpass, where the sky rms was higher due to the
presence of airglow lines, so we excluded these regions from the
automated search. 
The fraction of artificial emission line objects recovered
determines the completeness of our emission line search.
For the $0\farcs8$ FWHM, 18.4\AA\ FWHM sources, SExtractor finds
nearly all lines brighter than
$\sim$160 DN, or equivalently, to an emission line strength of $6 \times
10^{-18}$\flux\ for a source in the center of a slit.
Our results are summarized in Figure~\ref{fig:completeness} which
shows the completeness is higher for more compact lines.

Because of slit losses, the position of an object within the slit
affects its observed flux and hence impacts our ability to detect it.
Consequently, the effective area of our survey depends on the
intrinsic flux of the object: the faintest objects can only be
detected if they fall near the slit centre 
while intrinsically brighter objects can be detected closer to the slit edge
and hence over a larger area.  Figure~\ref{fig:slit_transmit} shows
how the effective area of our survey grows as the flux of the target
increases.  This plot includes the small, but measured, reduction
in search area caused by bright foreground objects that fall on a slit.
 The fluxes of sources that are only detectable 
exactly in the slit center -- where the survey area
tends to zero -- are $4.5 \times 10^{-18}$\flux, $6.2 \times
10^{-18}$\flux, and $8.0 \times 10^{-18}$\flux\ for our three fiducial
object sizes.  For objects that fall in the central two-thirds of a
slit (i.e. 1\farcs0), the area searched is 3.3895 square arcminutes
and the flux limit is $6 \times 10^{-18}$\flux, $8
\times 10^{-18}$\flux, and $10 \times 10^{-18}$\flux.  In
Section~\ref{sec:discussion} we will use the flux-dependent survey area
function shown in Figure~\ref{fig:slit_transmit} to predict expected
source counts.

\section{Object Identification}

The dividing line between faint sources and noise fluctuations 
becomes increasingly fuzzy as line flux decreases.
Based on the analysis of our simulations (see \S 2.3),
we believe most lines with 160 counts or more are real detections.
To  ensure that we did not miss objects,
we ran SExtractor (Bertin \& Arnouts 1996) with
the same detection criteria which were applied to the simulations 
in \S~2.3 and identified emission lines objectively.
Figure~\ref{fig:m1} shows the nine objects we found in the discovery spectra
above this count limit.

We extracted spectra for the nine emission-line objects, and
Table~\ref{tab:srclist} summarizes their measured line properties.
Notice that several line sources --  B5, B4b, A21, and C28 in
in order of increasing flux --  have line fluxes that if at z=5.7
would  imply \lya\ luminosities similar to  \lstarlya, where
$\lstarlya \equiv 3.26 \times 10^{42}$\ergsec. We adopt this
fiducial luminosity for the knee of the luminosity function 
based on the typical \lya\ luminosities  measured 
at z=3 (Steidel \et 2000) and at z= 4.9 (Ouchi 2003).
If \lstarlya\ does not evolve, an \lstarlya\ emitter at
z=5.7 has an observed flux of $1.09 \times 10^{-17}$\flux.  
Our faintest emission line
source, B5, would have a line luminosity of $0.77 \lstar$ if it were at
$z=5.7$.  For completeness, we remind the reader that
we do not know where in the slit a particular source fell so the
line fluxes are technically lower limits. 

%

Any viable high-redshift \lya\ emitter will have at least a factor of 5 to 10
continuum break just blueward of the \lya\ line due to attenuation from 
intergalactic gas (Madau 1995).  Figure~\ref{fig:m1} reveals
relatively bright continuum emission to both sides of the line in 
B5, D21b, and D22.  We reject these three objects as \lya\ candidates
on the basis of the absence of a continuum break; and we tentatively identify 
D22 as a blend of \Ha\ and [NII]~6584 on the basis of the line shape.  

For the other galaxies, the follow-up observations described in the
next section were required to determine their identity, but we draw
attention to a few interesting attributes derived from the discovery
spectra first.  For example, extremely high equivalent widths are often 
more easily fitted by the \lya\ line than any other emission line, and
we find C17 and D21 have extreme equivalent widths.  Specifically,
using the continuum blueward of the C17 line, we measure
observed equivalent width of 490\angs. No continuum is 
detected near D21, and we find the equivalent width must exceed 500\angs.
In addition to their high equivalent widths, objects
C17 and D21 are the only two candidates not spatially 
resolved in 0\farcs8 seeing.

\subsection{Follow-Up of Emission-Line Candidates}

\subsubsection{Follow-Up Spectroscopy}

To ensure accurate alignment of our follow-up mask on the desired
targets, we milled the follow-up mask using a modified version of
the mask design file from our search mask. 
Consequently, the relative positions of
set-up star boxes and those slit segments that contained the five
candidates of interest were preserved, allowing us to position our
follow-up mask's slits exactly on the same place in the sky as those
in our search mask.
To allow us to use higher dispersion and broader spectral coverage
for the follow-up observations without generating overlapping
spectra, slit  segments containing no \lya\ candidates were removed
from the mill file.   This procedure allowed us
to be confident that our targets were indeed in the slits of our
follow-up mask.

We obtained follow-up spectra of emission-line objects A21, B4b, B5,
D11, and D21 on 2002 November 9 with LRIS.  We inserted
the 831~${\rm mm}^{-1}$ grating blazed at 8200\AA\
to improve the spectral resolution on the red side
(0.93\AA\ pix$^{-1}$).  Blue light diverted by the D680 dichroic 
was dispersed by a 300~mm$^{-1}$ grism blazed at 5000\angs.
We removed the narrowband filter to increase the spectral
coverage from $\sim \lambda 6800$ to the atmospheric limit at $\sim 
\lambda 3200$ in the blue and $\lambda 6800$ to $\lambda 8700$ in
the red.\footnote{Exact coverage varies by object according to the
slitlit position.}  


Five hours of integration time were obtained
under thin cirrus clouds and 0\farcs8 seeing fwhm.  
We guided in the i-band to  keep the $\sim \lambda 8200$ 
emission centered in the slits.
Since there is no atmospheric dispersion compensator on LRIS,   
the targets are only in the slits on the blue side for a few hours.
The measured width
of the night sky lines was 4.7~pixels, or 4.4\angs.
The image processing steps were the same as those described in Section~2.2.
Figure~\ref{fig:m2} shows the emission line sources A21, D11, and D21 were 
redetected in our
follow-up spectra.  Object B4b was not re-detected.


\subsubsection{Imaging Follow-Up}

We mapped the positions of the emission line candidates onto a direct
image of the sky taken with LRIS.  Since we do not know the 
position of an object within the slit, the error perpendicular
to the slit direction is $\pm 0\farcs75$.  We computed only
the lowest order term for the coordinate transformation along
the slit, and the average error for a set of bright, test objects
was $0\farcs55$.  The coordinates on the LRIS image were tied
to a deep V-band image of CL0024V (Czoske \et 2001).  None of our emission 
line candidates were near bright stars or galaxies.  Object D21 lies along
a set of arclets near the cluster although the spectral follow-up
identified the line as \Ha\ emission from a foreground galaxy.  Although
our spectroscopic analysis yielded 3 emission lines which could be
\lya\ at z=5.7, we find a faint V-band object within 2\farcs0 of the
position of each. The small size of these galaxies is consistent
with either irregular galaxies at $z \sim 0.6$ or dwarf irregular
galaxies at $z \sim 0.24$
The faintest counterpart, related to C28, consists of two small clumps.
The brightest counterpart, related to B4b, appears to be a spiral galaxy
near, but not on, the slit. Deep $i$ or $z$
imaging of these fields would determine whether there is another
object closer to our predicted position.
Follow-up spectroscopy of each V-band counterpart would also quickly
determine if it is the emission-line source.

\subsection{Conclusions on Line Identifications}

We draw the following conclusions about the identity of the
emission-line objects in Table~1.

\begin{itemize}

\item{A21:  Several emission lines were detected in the follow-up
spectrum. We identify the discovery line as [OIII] $\lambda 5007$;
and the emitter is at $z = 0.656$. See Figure~6.
}

\item{B4b: Emission-line source not detected on follow-up mask.
}

\item{B5: Lack of a continuum break across the emission line
rules out \lya\ identification.
}

\item{C17: If the emitter is at z=5.7, the line luminosity
is brighter (and more rare) than the objects expected in our survey 
volume.  We tentatively identify a V band counterpart for this
object which, if correct, indicates the object is not at high
redshift.  We cannot rule  out a \lya\ identification at this
time however.
}

\item{C28: Marginal continuum detection to both sides of the line
leads us to suspect this object is not \lya.  Continuum detection
is too weak to rule out a \lya\ identification however.
}

\item{D11:  
Several emission lines were detected in follow-up spectrum. We
identify the discovery line as [OIII] $\lambda 5007$  at $z= 0.662$.
 See Figure~6.
}

\item{D21: This object lies near the cluster caustic. 
No continuum or line emission, aside from the single line
at $\lambda 8189.65$, was detected from D21 in the red follow-up
spectrum ($\lambda 7080 - \lambda 8970$).
The line profile has a Gaussian FWHM of 3.7\angs\ which 
rules out an identification as [OII] 3726, 29 at z=1.2.
However, a series of emission lines 
identified as [OIII] 4959, 5007, [OIII] 4364, and the Balmer series are
seen in the blue spectrum.
The high equivalent width line originally detected turns out to be  
\Ha emission at z=0.248. 
The 3-sigma upper limit on the [NII] 6583 line, which
is not detected, is $ < 1\%$ of the strength of the \Ha line.
  Our red spectrum covered [ArIII] 7135, HeI 6678, [SII] 6731,
  [NII] 6583, [OI] 6300, and He 5578. None of metal lines are strong
  in this obviously very metal poor galaxy; and the He I lines fell
  on strong night sky lines.
 }

\item{D21b: Strong continuum emission, but lack of continuum break rules
out \lya\ identification.
}

\item{D22:  Strong continuum detected, but lack of continuum break rules
out \lya\ identification. Second line redward of brightest line
presents correct offset for \Ha  and [NII] 6583 at a redshift $z \sim 0.24$.}

\end{itemize} 

Objects B4b, C17, and C28  are our only objects for which a 
\lya\ identification is plausible. 
Within the positional uncertainties, the position of each of these
emission-line sources matches the location of an object detected in 
a deep V-band  image.    In summary, we urge caution in the use
of single-line redshifts for high-redshift galaxies.  Many of
the faint emission-line objects that we found turn out to be
foreground galaxies upon more detailed inspection.  Our sample contains at most
three \lya\ emitters brighter than $7.8 \times 10^{-18}$\flux, and none of
these three candidates appears likely to be a genuine z=5.7 galaxy





\section{Discussion} \label{sec:discussion}

Since our survey sensitivity and search volume are well understood 
from simulations, our observations constrain the properties of the z=5.7 
population of Ly$\alpha$ emitters. 
The clean detection of objects B5 and C28  confirm that a \lya\  emission 
line with \lstar\ luminosity would easily be detected, and we are confident 
that galaxies as faint as 0.6\lstar\ would be detected $ > 85\%$ 
of the time.  The paucity of \lya\ emitters found excludes
a high density of bright galaxies with high \lya\ escape fractions.
We quantify this statement in \S 4.1 and \S 4.2.  In \S 4.3, we
discuss  whether bright,  \lya\ selected galaxies can maintain
the ionization of the intergalactic medium at z=5.7.


\subsection{Density of \lya\ Emitters}

To gain some insight into the properties of the high-redshift 
galaxy population, we parameterize the \lya\ emitting population
by a Schecter luminosity function $
\phi(L/\lstar) d(L/\lstar) = \phi^{*} (L/\lstar)^{\alpha}
e^{-L/L^*} d(L/L^{*})$, where the three parameters $\phi^{*}$,
\lstar, and $\alpha$ describe the number density, \lya\
luminosity of the bright-end turnover, and the faint-end slope.  
We expect our survey to find
\begin{equation}
N = \int_{L_{min}}^{L_{max}} \phi(L/\lstar) V(L/\lstar) \xi(L/\lstar)
d(L/\lstar)
\end{equation}
galaxies. The survey volume, $V$, and
the survey completeness, $\xi$, depend on luminosity because the
survey area and bandwidth decrease as the line flux approaches
the limiting flux of the survey (see \S 2.3).  We compute
$ V(L/\lstar)$ from the cosmological volume increment per unit
redshift, the accessible redshift interval (from Figure~1), and
the effective survey area (from Figure 4).  Figure~3 illustrates
the completeness function,  $\xi(L/\lstar)$.
Guided by the properties of the \lya\ population
at redshifts 3 and 4 (Steidel \et 2000; Ouchi \et 2003), we
adopt the following luminosity function --
$\phi^{*} = 0.0055$\mpc-3, $\lstar = 3.26 \times 10^{42}$\ergsec, and
$\alpha = -1.2$ (-1.6) --  and integrate down
to the limiting flux of the survey. For the no-evolution scenario,
we predict our survey should return 2.5 (2.4) \lya\ emitters per
field on average.


Using the above formalism,
we test the hypothesis that the luminosity function does not
evolve between redshift 3 and 5.7.  Suppose we observe many fields, 
we would expect to recover the mean -- i.e. 2 to 3 \lya\ emitters -- 
on average. 
Assuming (for the moment) the galaxies are randomly distributed in space
(i.e. Poisson statistics), the probability of obtaining 0 (or 1) object is
8.2\% (20\%).  The paucity of \lya\
emitters found in our field is therefore marginally consistent with
no-evolution in the \lya\ emitting galaxy population between redshift
3 and 5.7.


It is interesting to consider how strong lensing by the cluster 
CL0024+16 might affect the results of our survey.  
Galaxies seen through the cluster core will typically be magnified
by a factor $\sim 2$.  The cluster core subtends $\sim 18\% $ of the 
area covered by our mask, so we would expect 1 in 5 background galaxies 
to be lensed by a factor of two.  We find no statistically significant 
clustering among the  emission-line objects in our field.
Only two of our emission line
objects clearly have redshifts greater than the cluster redshift,
and one of these, A21, is projected well beyond the cluster core.
Given that we find no high redshift
galaxies, accounting for the presence of a lensing cluster in our 
field would strengthen our claim that the emission line galaxy
population does not undergo positive evolution from redshift 3 to 
redshift 5.7. Stronger magnification is only possible near
the caustics where our mask subtends an insignificant area.

Our results place strong constraints on the evolution of the
\lya -emitting population.
For any model of the luminosity function described by
$\phi^{*}$, \lstar, and $\alpha$, we can calculate
the number of \lya -emitters our survey would detect.  
\fig~\ref{fig:constraints}a shows 
the expected yield of \lya\ emitters per field over a grid 
spanning a factor of three evolution in \lstar\ and $\phi^{*}$.
For each of these models, we computed the probability of our field 
yielding  zero \lya\ detections.
The leftmost dashed line in \fig~\ref{fig:constraints}b illustrates
the 99\% Poisson confidence contour for our null result.
We find that in the limit of no density evolution, the value of
\lstarlya\ appears unlikely to brighten by more than a factor 1.8 
between z=3 and z=5.7.  Similarly, in the limit of no luminosity evolution, 
the number density of \lya\ emitters does not increase by more than
a factor of 2.1 between z=3 and z=6.  
The number of predicted objects
is not sensitive to the faint-end slope of the luminosity
function because our observations reach only 0.6\lstar.\footnote{
  \fig~\ref{fig:constraints}a and
  \fig~\ref{fig:constraints}b use a faint-end slope of
  $\alpha = -1.2$. 
  For a steeper faint-end slope with $\alpha = -1.6$, the 
  allowed brightening of \lstar\ is slightly smaller (1.7 for zero \lya
  emitters detected) while the allowed increase in $\phi^{*}$ is slightly 
  larger (2.2 for zero objects detected).}
The other curves show the
upper limits had we detected 1, 2, or 3 \lya\ emitters.  Even if
a similar observation of another field yielded 2 \lya\ detections,
we would conclude that between z=3 and z=6 \lstar\ brightens by less 
than a factor 3 and the comoving density of \lya\ emitters grows by less 
than a factor of 3.75.  
Our results rule out strong positive
evolution in the \lya - emitting population from z=3 to z=6.

\subsection{Impact of Galaxy-Galaxy Clustering on Results}

The true variance in galaxy counts $N$ about the mean $\mu$ will
be larger than the Poisson variance, $\delta^2 = <(N - \mu)^2> = \mu$,
by an amount reflecting the clustering strength, $\mu^2 \xi(r)$ 
(e.g. Peebles 1993).  The correlation function, $ \xi(r)$,  
has not yet been measured for high-redshift \lya\ emitters, but
we can illustrate how much clustering would be required to significantly
change our conclusions.
If clustering reduces the number of \lya\ emitters in our field
from 2.5 to 0, then $\delta \rho / \bar{\rho} = 1$ and $\xi(r) = 1$.  
The variance of the counts increases to $\delta = \sqrt{\mu + 
\mu^2 \xi(r)}  = 2.0$, which is not much larger than the value of
$\delta = 1.6$ expected for a random source distribution.  Yet
this estimate implies a higher degree of clustering than the
Steidel \et (1999) $\delta \rho / \bar{\rho} \approx 0.25$
estimate for \lya\ emitters at $z \sim 3$.
It seems unlikely, therefore, that clustering of \lya\ 
emitters could account for the paucity of \lya\ emitters found by our
survey; and our conclusion - that \lstar\ \lya\ emitters cannot
increase much in luminosity or increase in number density between $z \sim 3$
and $z \sim 5.7$ -- holds even if the population is fairly strongly
clustered.

To place further constraints on the potential impact of clustering,
it is interesting to ask whether our number density is consistent with 
the yield of high-redshift galaxies from other \lya\ surveys.  
For a non-evolving \lya\ population, we expected to recover only
a couple \lya\ emitters in our pilot survey, and we expected them to 
have sub-\lstar\ luminosities.   Narrowband imaging surveys, in contrast,
cover larger areas but recover only objects brighter than roughly 2\lstar.
The no-evolution model implies that a survey must, on average, cover 21
square arcminutes on the sky to return one 2.0\lstar\ \lya\ emitter
from a 103~\AA\ wide bandpass at z=5.7.  This area corresponds to 
a survey volume $\sim 4500$ Mpc$^3$. The LALA group has three
secure identifications of $z=5.7$ \lya\ emitters which corresponds to
$2.5 \times 10^4$~Mpc$^3$  per 2.5\lstar\ \lya\ emitter.  For our
adopted cosmology, the volume
per emitter for the Hu \et narrowband imaging is smaller at roughly 
6900 ~Mpc$^3$. The \lya search at Subaru confirmed a bright {\bf $z = 5.7$}
emitter ($F = 4.1 \times 10^{-17}$\flux; Taniguchi \et 2003) after
searching about $3.4 \times 10^4$~Mpc$^3$.
Considering the uncertainties inherent in gross comparisons among
these surveys, our limits on the number density of \lya\ emitters 
appear to be consistent with the other results.  It remains plausible
then that the \lya\ emitting population could maintain the ionization
of the IGM at $z = 5.7$ and be related to the population responsible
for cosmic reionization.

Lehnert \& Bremer (2003) selected galaxies between z=4.8 to z=5.8.
using the broadband drop-out technique and spectroscopically confirmed two 
\lya\ emitters in the 8200\angs\ atmospheric window.  
Their survey volume in the z=5.697 to z=5.710 window is only
$\sim 9.72 \times 10^3$~Mpc$^3$, but both objects found 
are bright sources typical of those found by narrowband-selection
(BDF 1:10 $F=2.45 \times 10^{-17}$; BDF 2:19 $F=2.51 \times 10^{-17}$).\footnote{They also find a much fainter source just redward of the window 
which has an  observed equivalent width of 40\angs\ and would be
difficult to recover in a narrow-band imaging survey.
The object, BDF 1:19, has flux $3.09 \times 10^{-18}$\flux - just below the 
sensitivity limit of our survey.}
The implied number density of $L \sgreat\ \lstar$ \lya\ emitters in the
field is roughly 1 per 5000~Mpc$^3$.  
Lehnert \& Bremer (2003) conclude that their objects cannot maintain
the ionization of the IGM and suggest that lower luminosity galaxies 
contribute to the ionization of the IGM.  Since some of the emission
line galaxies that we detected are too faint in the continuum to be
selected by the Lehnert \& Bremer survey, their result may
be consistent with our conclusion.
To date then, the different high redshift galaxy 
surveys seem to yield mutually plausible
constraints on the high redshift galaxy population. Consequently, 
we cannot yet use the variance in \lya\ emitting galaxies among surveys
to empirically estimate their clustering properties.

\subsection{Are There Enough High-Redshift Galaxies to Ionize the 
Intergalactic Medium?}

Recent observations of high redshift quasars pin 
the redshift of the final reionization epoch near $z \sim 6$ (Becker \et 2001;
Djorgovski \et 2001), so the formation rate of stars and quasars must
remain high enough following the end of Reionization to at least offset 
the recombination rate of the intergalactic gas.
The identity of the objects that maintain the ionization of the
intergalactic medium at $z=5.7$ remains an open question.  While quasars 
dominate the production of intergalactic Lyman continuum radiation 
by $z=0$, at $z \sim 3$ more ionizing flux escapes from Lyman-break-selected 
galaxies than from quasars (Madau \et 1999; Steidel \et al 2001).  
The contribution from quasars at $z \sim 5.7$ appears to fall far
short of that needed to maintain IGM ionization (Fan \et 2001).
It is interesting to examine whether the bright \lya\ emitting 
population is a plausible source of the required ionizing luminosity.

The time from redshift z=5.7 to the reported completion 
of Reionization at z=6 (Becker \et 2001; Djorgovski \et 2001) is
only $152 h_{0.7}^{-1}$~Myr.  Hence  the galaxy population responsible
for Reionization may still be present at $z=5.7$.  A hydrogen
atom in the IGM consumes several ionizing  photons in this short
period. The recombination time at z=5.7,
averaged over a clumpy IGM,  is 
$\bar{t}_{rec} = 64~{\rm Myr~} (0.041 / (\Omega_b h_{70}^2)) C_{30}^{-1}$,
where $C_{30}$ is the ionized hydrogen clumping factor  in units of 
$C \equiv <n_{HII}^2> / \bar{n}_{HII}^2  = 30$ (Madau \et 1999).  
To raise the filling factor, $f_{HII}$, of ionized regions in the IGM
to unity, the number of ionizing photons escaping from galaxies over
the recombination timescale must equal or exceed the baryon density.  
Following Madau \et (1999), the critical production rate of 
ionizing photons is $\dot{N}_H = \bar{n}_H(0) / \bar{t}_{rec}(z)$
per unit comoving volume, which
we can write as
\begin{equation}
\dot{N}_{H} = 10^{51.34} {\rm ~s}^{-1} 
{\rm ~Mpc}^{-3}~ C_{30} \left( \frac{1 + z}{ 6.7} \right)^3
\left(  \frac{\Omega_b h_{70}^2}{0.041} \right)^2.
\end{equation} 
For an initial stellar mass function $dM/dN \propto M^{-2.35}$ (i.e. Salpeter 
IMF), continuous star formation produces ionizing photons at a rate of 
$\dot{N}_H = 10^{53.4}$~s$^{-1}$ per solar  mass of 1.0\msun\ to 100\msun 
stars formed in 1 year (Leitherer \et 1999).  Including the mass
of 0.1 to 1.0\msun\ stars formed by this IMF, the minimum cosmic
star formation rate required to keep the universe ionized at z=5.7 
in the absence of any ionizing flux from quasars is
\begin{equation}
 \dot{\rho}_{*} = 0.0223 \msunyr {\rm Mpc}^{-3} C_{30} \left( \frac{1 + z}
{6.7} \right)^3 \left( \frac{\Omega_b h_{70}^2}{0.041} \right)^2 f_{LyC}^{-1},
\end{equation}
where $f_{LyC}$ is the fraction of Lyman continuum photons escaping from 
galaxies. Theoretical simulations predict that the escape fraction, $f_{LyC}$, 
will become large when a starburst-driven wind blows out of a galactic
disk (Fujita et al. 2003). In nearby starburst galaxies, measured values
of $f_{LyC}$ are less than 15\% (Leitherer \et 1995; Hurwitz \et 1997).
Some high-redshift galaxies  have higher escape fractions
(Steidel \et 2001),  but no clear consensus about escape fractions
from  high-redshift starbursts has been reached (Fern\'{a}ndez-Soto \et 2003). 
Although the most appropriate value of $f_{LyC}$ for high-redshift
galaxies clearly remains to be determined, values up to unity remain
plausible for now.

The \lya\ line is a notoriously poor measure of the galactic star
formation rate.  The large scattering cross section in the line 
produces a high probability that a \lya\ photon will be absorbed by a 
dust grain.    The effective optical depth in the line can
be dramatically reduced, however, by bulk motions in the ISM which
Doppler shift the line photons away from the resonance.  
Observations  of the \lya\ emission/absorption profile show wide
variety among even starburst galaxies (Shapley \et 2003), and models
suggest the emission line flux depends as much on the
evolutionary state of the starburst wind as on the absolute value of 
the star formation rate (SFR) (Tenorio-Tagle \et 1999).  For purposes of 
illustration,
we simply parameterize the line luminosity in terms of the fraction
of \lya\ photons that escape the galaxy $f_{Ly\alpha}$.\footnote{
   Note that $f_{Ly\alpha}$ is technically the escape fraction
   of \lya\ photons relative to the escape fraction of \Ha photons, but
   we implicity assume all the \Ha\ photons escape.}
For a Salpeter stellar IMF extending from 0.1\msun\ to 100\msun\,
a star formation rate of 1.0 \msunyr\ yields an \Ha emission line luminosity 
of $1.26 \times 10^{41}$~ergs~s$^{-1}$ (Kennicutt 1998). 
Using the Case~B ratio of the H recombination lines at low density
(Ferland \& Osterbrock 1985),  we find
$L_{Ly\alpha} = 8.32 f_{Ly\alpha} L_{H\alpha}$. 
Substituting for the star formation rate
in Equation 3, we derive the critical (comoving) luminosity density of \lya\ 
emission,
\begin{equation}
\L_{Ly\alpha}  = 2.34 \times 10^{40} {\rm ~erg~s}^{-1} {\rm ~Mpc}^{-3}~ 
\left( \frac{f_{Ly\alpha}} {f_{LyC}} \right)
C_{30} 
\left( \frac{1 + z}{6.7} \right)^3 \left( \frac{\Omega_b h_{70}^2}{0.041} \right)^2,
\end{equation}

required to maintain the ionization of the IGM at z=5.7.

We may use Equation~4 to explore whether the population of z=5.7 \lya\
emitters can make a significant contribution to the required ionizing
luminosity.  For our fiducial model ($\phi_{*} = 0.0055$~Mpc$^{-3}$,
 $\lstarlya = 3.26 \times 10^{42}$, and $\alpha = -1.2$), the \lya\ 
luminosity produced by galaxies brighter than 0.6\lstar\ is 
$\L = 0.0028 \lstar$~Mpc$^{-3}$. This population can only maintain the
ionization of the IGM if $f_{Ly\alpha}$ is small (relative to $f_{LyC}$) - 
i.e. most of the emitted \lya\ is absorbed and therefore not detected.
Using equation~3, our limits on the number density of 
galaxies brighter than 0.6\lstarlya\  imply the escape fraction
of \lya\ photons would have to be $f_{Ly\alpha} \le\ 0.39 f_{LyC}$
if these relatively bright galaxies were to maintain the ionization
of the IGM.  

Since the optical depth in the
\lya\ line is much higher than the optical depth at the Lyman limit,
it is possible for a large fraction of the Lyman continuum photons to escape
while most \lya\ photons are absorbed by dust after many scatterings.
A \lya\ escape fraction  $f_{Ly\alpha} \sim 0.4$ also seems plausible
when the properties of the known $z \sim 6$ galaxies are considered.
For SSA22-HCM1 (z=5.74), Hu \et (1999) derive a SFR 
from the ultraviolet continuum that is twice
as large as that derived from the \lya\ line. (Using the flux
to SFR scale adopted in this paper, we get a SFR from the continuum
that is six times larger than that estimated from the line flux, so a
value of $f_{Ly\alpha}$ in the range of 0.2 to 0.5 is at least plausible.)
If the number of luminous \lya\ emitters at z=5.7 is similar
to our inferred upper limit on their number density and the \lya\ escape 
fraction $\sles\ 40\%$, then relatively bright \lya\ emitters with 
$L \sim\ \lstarlya\ $ can maintain the ionization of the IGM at $z = 5.7$.  
A significant deficit of \lstar\ \lya\ emitters relative to 
this no-evolution model would suggest that another population of
galaxies, perhaps large numbers of dwarf galaxies, must be present to 
maintain the ionization of the IGM.  That measurement  should be possible
soon.

Continuum-break-selected surveys also find a paucity of galaxies
at high redshift.
In particular, Lehnert \& Bremer (2003) conclude that the bulk of the ionizing
flux that reionized the universe came from faint galaxies with 
$M_{AB}(1700 {\rm \AA}) > -21$.  They
find a lower density of $z \sim 5.3$ r-band dropouts than expected from 
no-evolution predictions tied to the lyman-break-selected galaxy
population at z=3 and 4.
The ultraviolet luminosity density from their $z \sim 5.3$ drop-outs 
is considerably less than that required to ionize their survey volume.  
Drop-out searches complement the \lya\ searches because
the inherent biases are quite different.  (The dropout
technique yields some false-positive detections, 
but the \lya\ surveys may miss galaxies entirely.)  Constraints from
both techniques will probably be required to pin down the population
producing the ionizing background at $z \sim\ 6$.

\section{Conclusions and Outlook}

We have shown that the multi-slit windows technique 
provides a very sensitive means of searching for 
emission-line galaxies.  It probes significantly 
larger volumes than surveys of cluster caustics but reaches
sub-\lstar objects which are largely inaccessible to
narrow-band imaging surveys at high redshift.  In our pilot survey,
most of the emission-line sources were found to be foreground 
galaxies.  It is  possible that the \lya\ emitting population at 
z=5.7 is described by the same \lstar\ and $\phi_{*}$ as the
\lya -emitting population at $z \sim 3$ to 4, and we recovered no \lya\ 
emitters simply because the richness of our field was poor compared
to an average field. However, the paucity of \lya\ emitters
in our field does strongly rule out significant positive evolution
in the luminosity function. Neither the characteristic luminosity nor 
number density of \lya\ emitters can increase much between z=3 and z=5.7 
-- a period long enough, $1.1 h^{-1}_{0.7}$~Gyr, to allow several cycles
of bursting activity and fading (Sawicki \& Yee 1998; Shapley \et 2001).

Since the recombination time of the intergalactic gas is shorter
at higher redshift, the paucity of \lya\ emitters at $z \sim 5.7$ 
raises the question of whether the \lya -selected population 
can keep the IGM ionized at $z = 5.7$.  In the no-evolution scenario,
in order for the inferred star formation rate to be high enough
to maintain the ionization of the intergalactic medium, the
fraction of \lya\ emission escaping must be less than 40\%. 
This  average \lya\ escape fraction does not appear to be unusual.
Indeed, only 25\% of starburst galaxies at z=3 have \lya\
in emission (Shapley \et 2003).  
Our constraint is strictly an upper limit, however.  
Additional observations will either recover \lya\ emitters in number or 
push this upper limit low enough to require
another population of galaxies (e.g. perhaps dwarf galaxies)
 to maintain ionization.
The existence of another  population of dwarf starbursts is suggested by the
detection of a z=5.7 \lya\ emitter with de-lensed flux of only $2 \times
10^{-18}$\flux\ in a search volume an order of magnitude smaller than
that of our survey (Ellis \et 2001). 

Future surveys with the multislit windows technique are expected
to produce about a dozen high-redshift galaxies per pointing.
The gains will come from a combination of improved sensitivity,
the larger field of view provided by new cameras, and multi-night
observing campaigns.  
Dispersing the spectra to a rest-frame velocity dispersion of 200\kms,
which is about $\sim 5.5$\angs, will reduce the sky noise by another
factor of 3 to 4.  The throughput can probably be doubled by using
a more efficient dispersing element.  It should be possible, therefore,
to reach fluxes a factor of 2.5 to 3 times fainter than our limit of 
$6.2 \times 10^{-18}$\flux\ in a single night of observing time.
Although the cost of higher dispersion would normally be 
reduced area, the new generation of large format CCD arrays 
more than makes up for the decrease in survey area.  

Atmospheric windows at longer wavelength allow the technique to
be extended to higher redshifts.  However, even in the next window
at 9200\angs\ (z = 6.5), the exposure time required to reach
a given luminosity is nearly a factor of two higher than that
required at z=5.7 (in the 8200\angs\ window).  We would expect
a large multislit windows survey in the 8200\angs\ window 
to yield a secure measurement of the number density of
dwarf (i.e. sub \lstar) \lya\ emitters at z=5.7.

\acknowledgements{These observations would not have been possible
without the narrowband filter kindly lent by Ester Hu. We thank David Crampton 
for his encouragement and useful discussions.  CLM and MJS recognize and 
acknowledge the very significant cultural role and reverence that the summit 
of Mauna Kea has always had within the indigenous Hawaiian community.  We are 
most fortunate to have the opportunity to conduct observations from this 
mountain. This work was funded in part by grants from the David and Lucile
Packard Foundation (CLM) and the Alfred P. Sloan Foundation (CLM).}

\clearpage

\begin{table} 
\caption{Properties of Emission-Line Sources}
\begin{tabular}{lllllll}
\hline
Object ID&  FWHM      & FWHM   & Line Flux & Line Flux & EW$_{\rm o}$  & Line ID   \\
(MS-0h17)&  (\arcsec) & (\angs)& (Counts)  & ($10^{-17}$\flux)             & (\angs) & (redshift)\\
\hline
\hline
A21      &  1.47      & 18.0   & $340 \pm 50$  & $1.30 \pm 0.2$   & 28      & [OIII], z=0.656  \\
B4b      &  1.48      & 15.1   & $290 \pm 29$  & $1.1 \pm 0.1 $   & $>80$   & Unlikey \lya\ Line \\
B5       &  1.40      & 10.0   & $220 \pm 50$  & $0.84 \pm 0.2$   &  9      & Not \lya\ \\
C17      &  1.10      & 17.2   & $1200 \pm 44$ & $4.80 \pm 0.2$ & 490     & Unlikely \lya\ Line   \\
C28      &  1.29      &	19.5   & $430 \pm 56$  & $1.7 \pm 0.1$   & 80      & Unlikely \lya\ Line  \\
D11      &  1.08      & 16.5   & $640 \pm 38$  & $2.5 \pm 0.2 $	 & 47      & [OIII], z=0.662  \\
D21      &  0.86      & 17.7   & $6280 \pm 66$ & $24.4 \pm 0.3$	 & $> 500$ & \Ha at z=0.248      \\
D21b     &  1.38      & 20.8   & $700 \pm 90$  & $2.7 \pm 0.4$ & 32      & Not \lya\  \\
D22      &  1.16      &	20.4   & $2760 \pm 110$ & $10.7 \pm 0.4$   & 40      & Likely \Ha +[NII]  \\
\hline
\end{tabular}

Table Notes -- 

(1) Object name on Martin \& Sawicki  mask for field at 0h 17\deg.

(2) Extent of emission line along the slit.

(3) Observed line width.  All lines are unresolved.

(4) Counts in the emission line.

(5) Line flux derived assuming the average filter transmission over the bandpass.

(6) Emission-line equivalent width.

\label{tab:srclist} \end{table}

\clearpage


 {\par\centering

\newpage

\begin{figure} 
       {\psfig{file=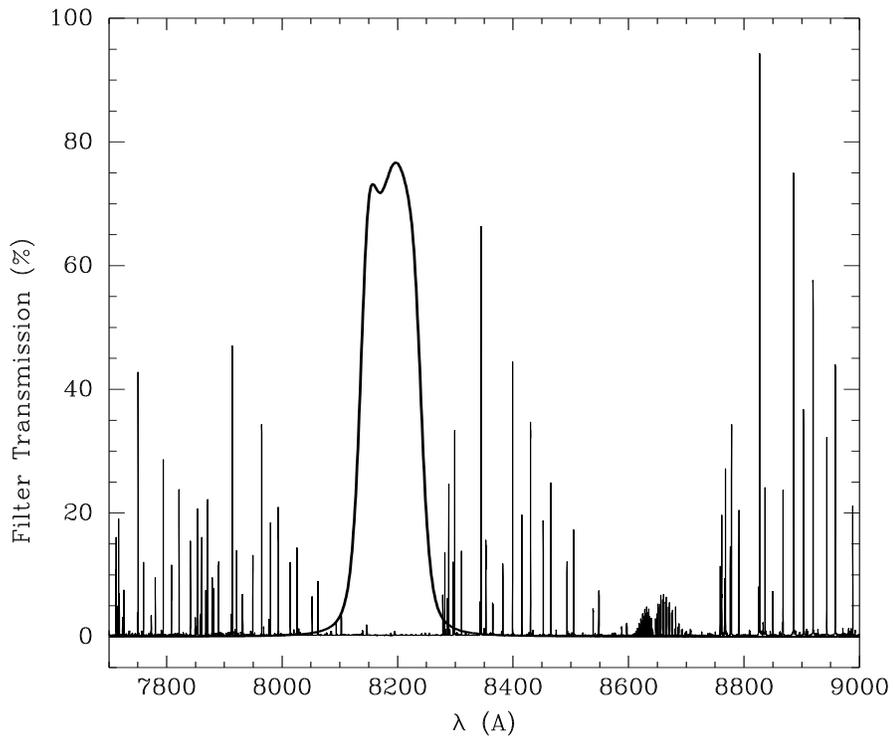,angle=-90,height=5.0in} \hfill}
\caption{Relative intensity of the Mauna Kea night sky (courtesy of
Tom Slanger). The transmission of filter NB8185 (kindly loaned by Esther Hu) 
illustrates the atmospheric window probed by our survey.}
\label{fig:skywindow} \end{figure}

\newpage

\begin{figure} 
       {\psfig{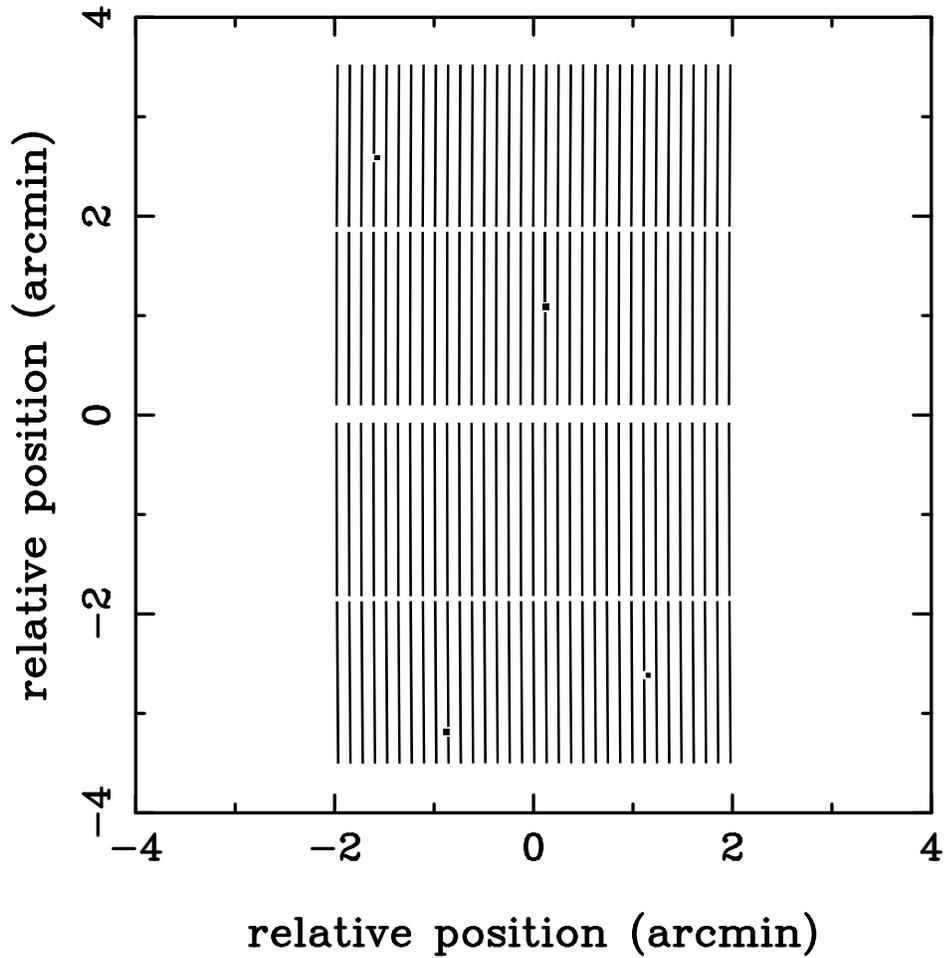} \hfill}
\caption{Slitmask layout for the search mask.  The mask consisted of 33 
longslits interrupted by three bars needed for structural rigidity.    
Small squares are holes for alignment stars used for positioning the
mask on the sky.  Dispersion direction is left to right.
}
\label{fig:mask} \end{figure}

\newpage

\begin{figure} 
       {\psfig{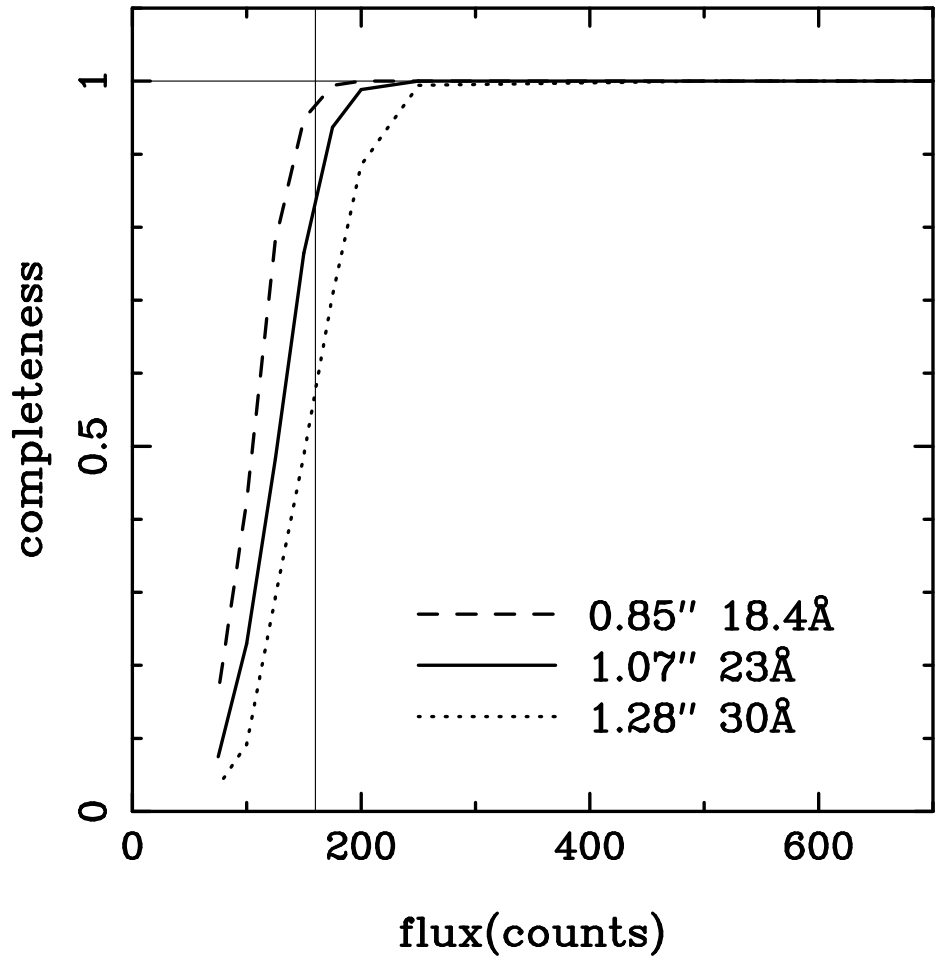} \hfill}
\caption{Survey sensitivity.  The fraction of simulated objects recovered by 
the automated detection routine for the three combinations of object
size parameters. We define our completeness limit at 160 counts,
which, in the absence of slit losses, corresponds to 6$\times$$10^{-18}$
erg s$^{-1}$ cm$^{-2}$.
}
\label{fig:completeness} \end{figure}

\newpage

\begin{figure} 
       {\psfig{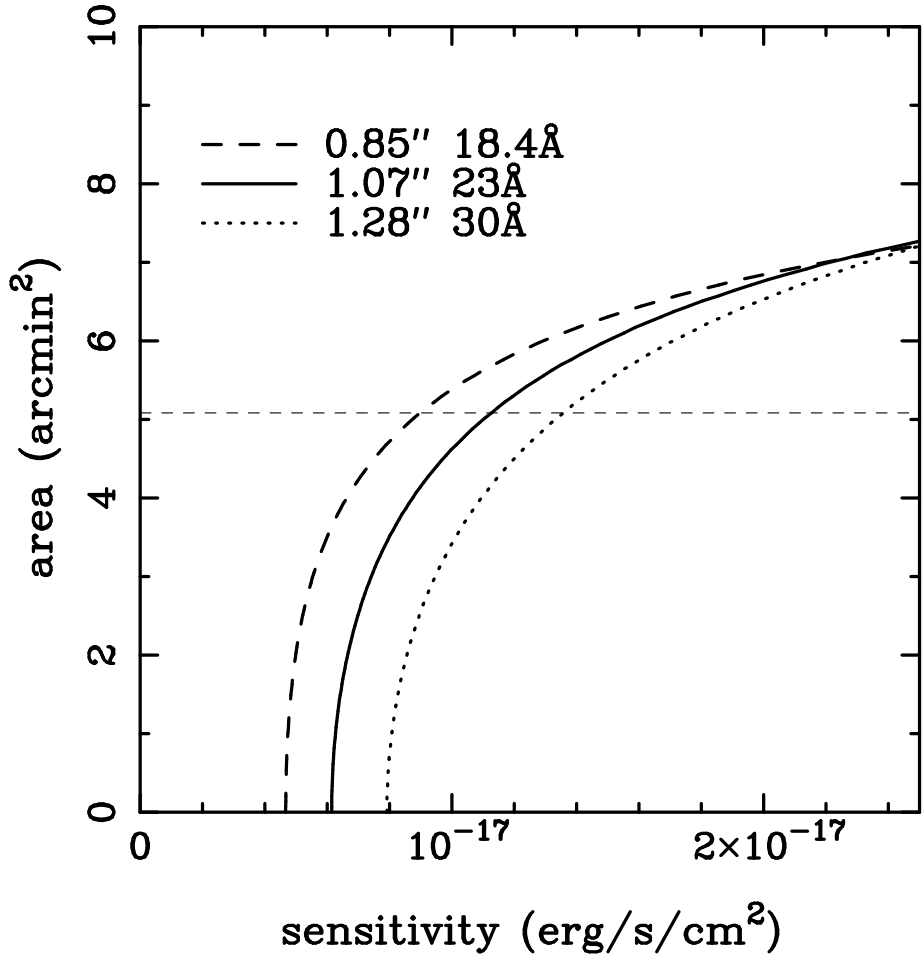} \hfill}
\caption{Our survey area as a function of emission line flux. The
horizontal line denotes the geometric angle subtended by the slitlets.
Slit losses 
reduce the transmitted flux, and the faintest emission lines are only detected
when the objects land in the center of the slit. 
}
\label{fig:slit_transmit} \end{figure}

\newpage

\begin{figure} 
\caption{Emission-line candidates found by the automated
search. The arrows denote 5\asec and
100\angs\ in the spatial and spectral directions respectively.
Wavelength increases from left to right.}
\label{fig:m1} \end{figure}

\newpage

\begin{figure} 
       {\psfig{file=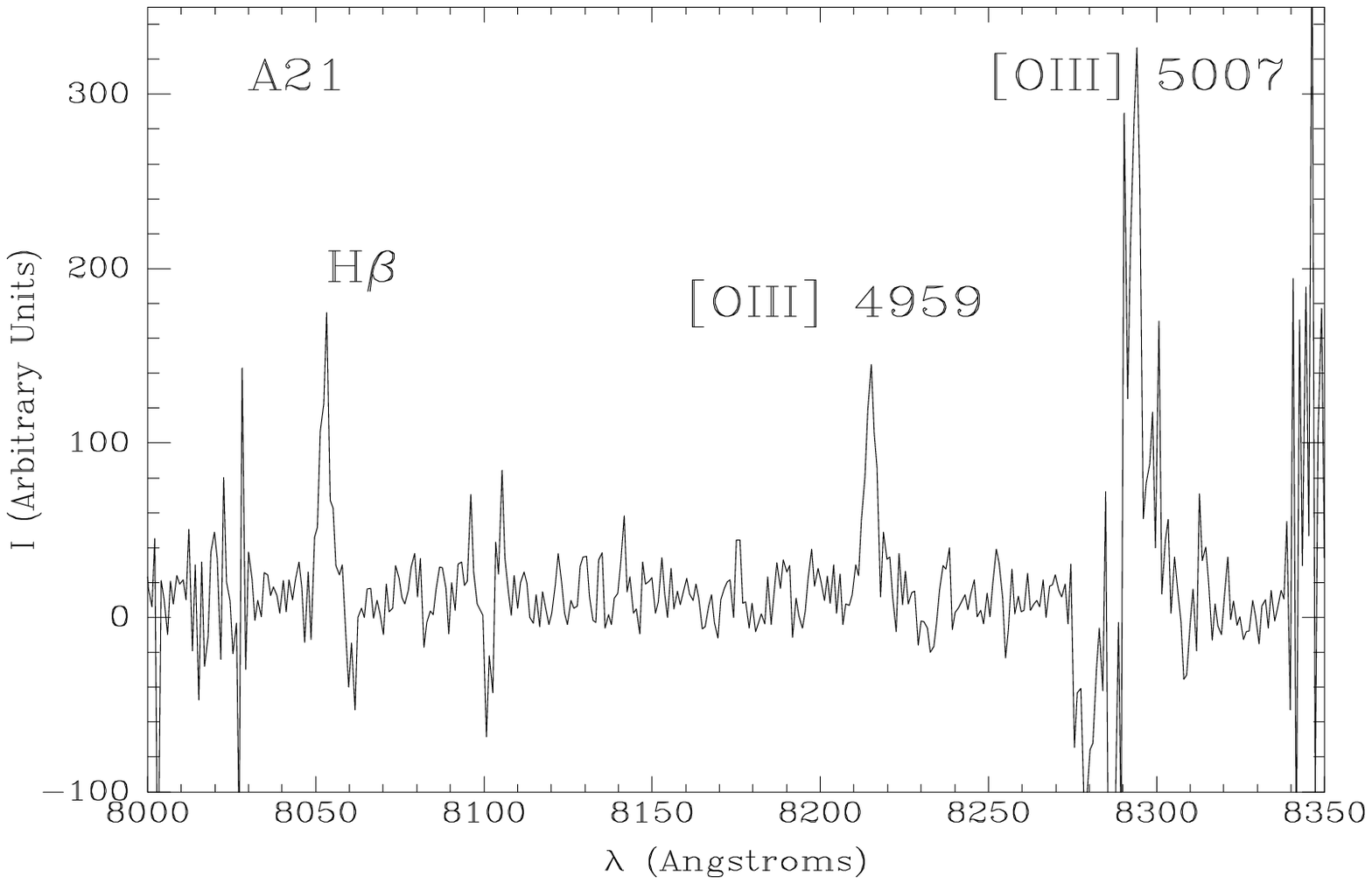,angle=0,height=2.5in} 
        \psfig{file=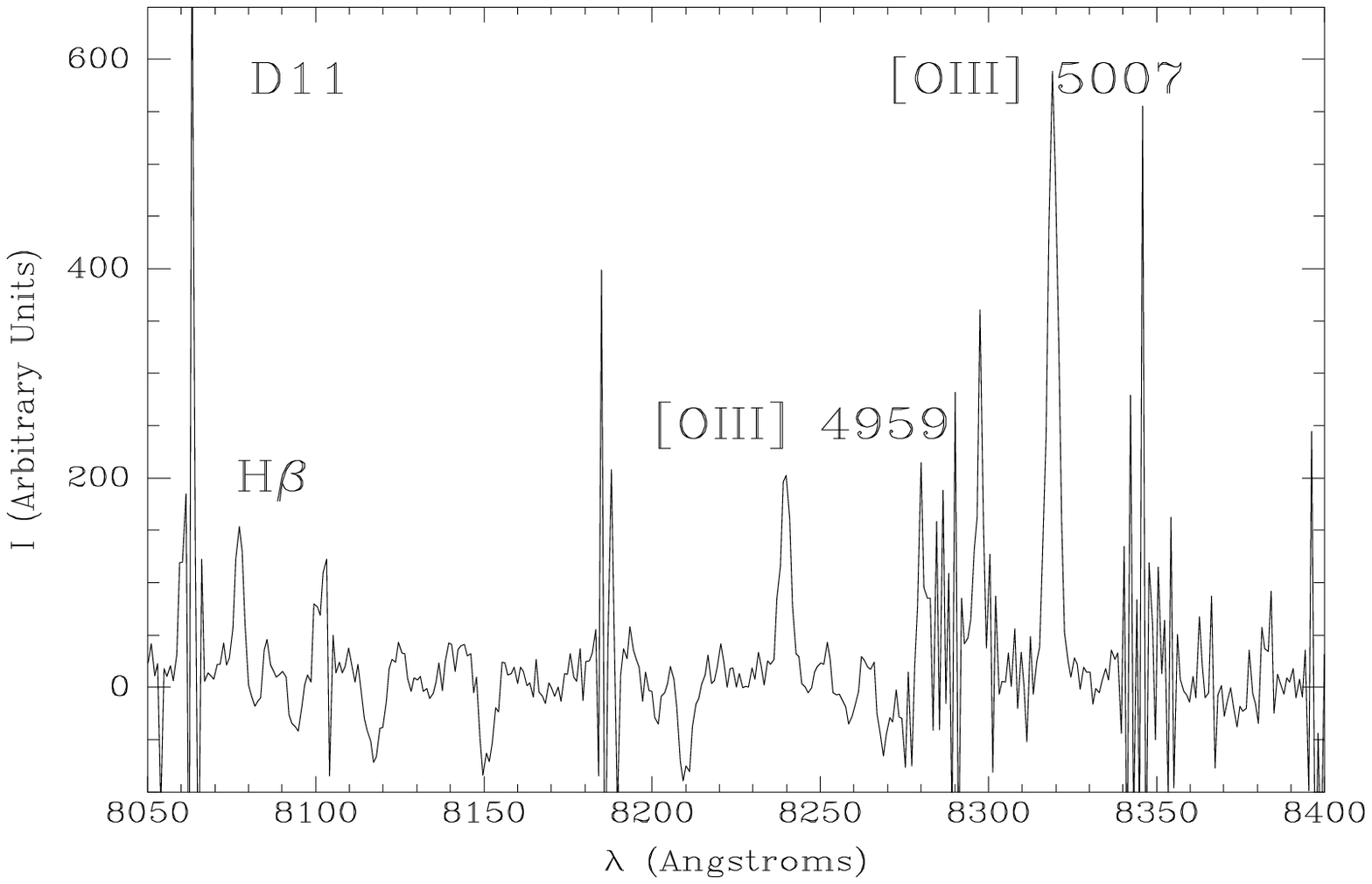,angle=0,height=2.5in} 
        \psfig{file=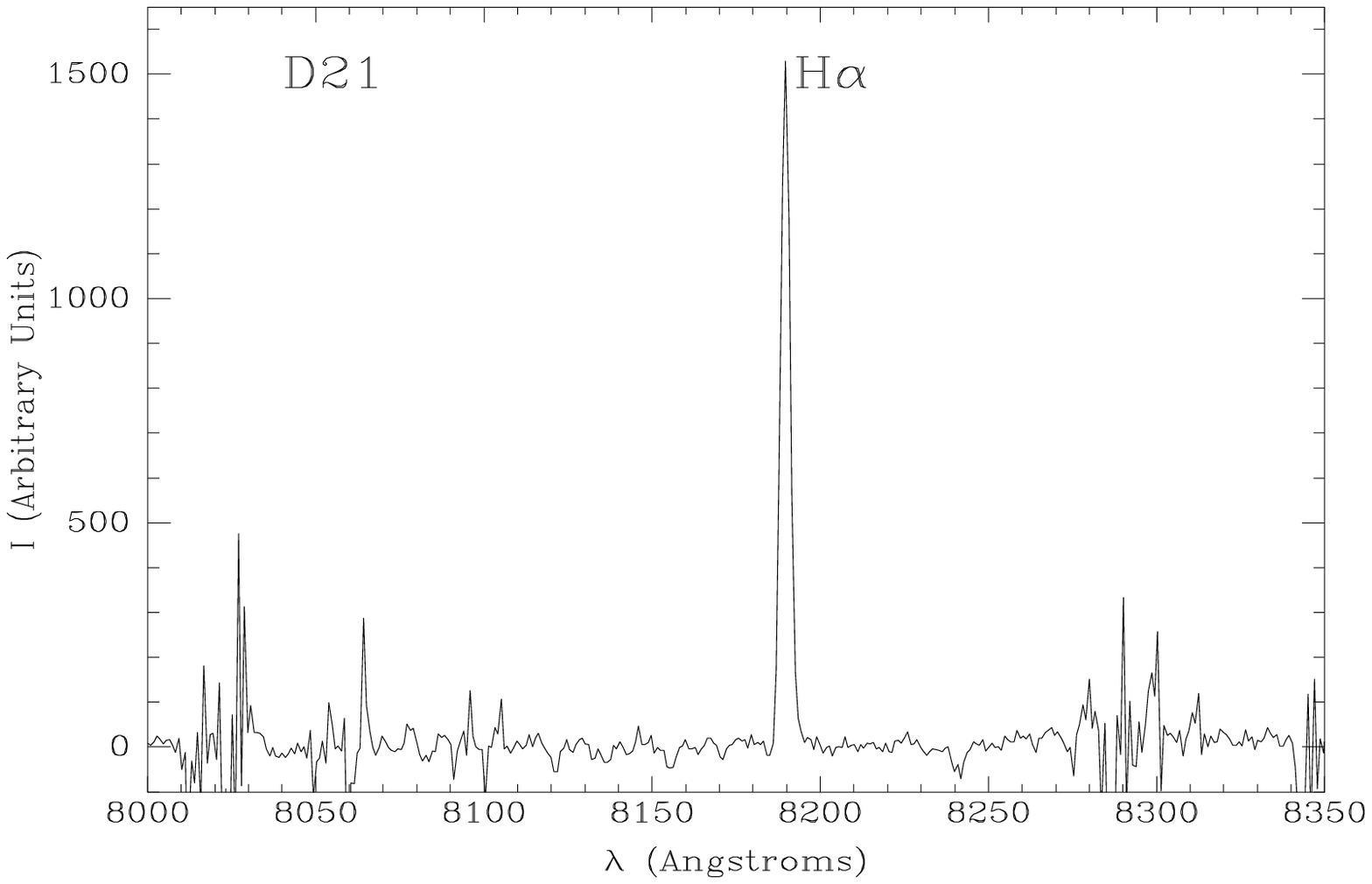,angle=0,height=2.5in} 
	 \hfill}
\caption{Follow-up spectra for objects A21, D11, and D21.  
None of the confirmed line identifications are \lya.}
\label{fig:m2} \end{figure}

\newpage

\begin{figure}
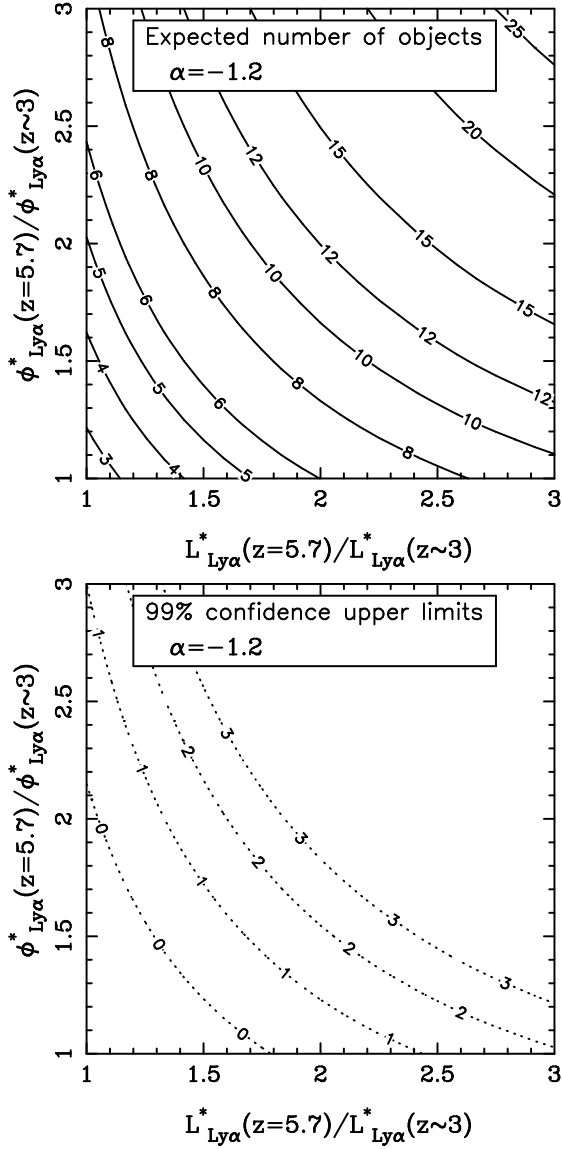
 
       {\psfig{file=martinc.fig7a.ps,angle=0,height=3.0in}
         \psfig{file=martinc.fig7b.ps,angle=0,height=3.0in} \hfill}

\caption{(a) Predicted survey yield for various evolution models of the \lya\
luminosity function.  The axes indicate the amount of evolution in the
number density (vertical axis) or luminosity density (horizontal axis)
relative to our baseline luminosity function
( $\lstarlya (z=3) \equiv 3.26 \times 10^{42}$~ergs~s$^{-1}$,
 $\phi^*_{Ly\alpha} (z=3) = 0.0055$~Mpc$^{-3}$). The faint end slope of 
$\alpha$=-1.2 was assumed for the luminosity function, but the results do 
not depend significantly on the assumed $\alpha$. 
 For a Ly$\alpha$ luminosity function
 that does not evolve between $z$$\sim$$3$ and $z$=5.7,
we expect to recover $\sim$2.5 $z$=5.7 Ly$\alpha$ emitters on average. 
(b)Constraints on the evolution of the Ly$\alpha$ luminosity 
function  from $z$$\sim$3 to $z$=5.7.   The leftmost curve ---
labelled ``0'' --- shows the upper limit (Poisson 99\% confidence) on
the amount of luminosity and density evolution in the Ly$\alpha$ 
luminosity function allowed by the null result of our survey.
The area of parameter space to the right of the curve is ruled out by
our lack of Ly$\alpha$ detections.  The other curves 
show the upper limits had we detected 1, 2, or 3 Ly$\alpha$ emitters.
Evidently between redshift 3 and 6,
\lstarlya\ does not brighten by more than a factor of 1.8 and
$\phi^*_{Ly\alpha}$ does not increase by more than a factor of 2.1.} 
\label{fig:constraints} \end{figure}

\newpage






\begin{references}

\reference{} Ajiki, M. \et 2002, \apj, 576, L25 

\reference{} Bertin, E. \& Arnouts, S. 1996, A\&AS, 117, 393 


\reference{} Becker, R. H. \et 2001, AJ, 122, 285

\reference{} Bunker, A. J. \et 2003, \mn, 342, L47

\reference{} Chapman, S. C., Blain, A. W., Ivison, R. J., \& Smail, I. 2003, Nature, April 17. 

\reference{} Crampton, D. \& Lilly, S. 1999, ASP Conf. Series, 191, p. 229

\reference{} Czoske, O. \et 2001, \aa, 372, 391 

\reference{} Djorgovski, S. G. \et 2001, \apj, 560, 5

\reference{} Ellis, R. \et 2001, \apj, 560, 119 

\reference{} Fan, X. \et 2001, \aj, 122, 2833 

\apj, 569, L65

\reference{} Ferland, G. J. \& Osterbrock, D. E. 1985, \apj, 289, 1985

\reference{} Fern\'{a}ndez-Soto, A. F., Lanzetta, K. M., \& Chen, H. W. 2003,
astro-ph/0303286

\reference{} Fujita, A., Martin, C. L., Mac Low, M.-M., \& Abel, T. 2003, 
submitted to \apj

\reference{} Hurwitz, M., Jelinsky, P., \& Van Dyke Dixon, W. 1997, \apj, 281, L31

\reference{} Kennicutt, R. C. 1998, \apj, 498, 541 

\reference{} Kodaira, K. \et 2003, PASJ, 55, 17 

\reference{} Kogut, A. \et 2003, astro-ph/0302213

\reference{} Hu, E. \et 2002, \apj, 568, L75 

\reference{} Hu, E. M., McMahon, R. G., \& Cowie, L. L. 1999, \apj, 522, L9


\reference{} Lehnert, M. \& Bremer 2003, astro-ph/0212431 

\reference{} Leitherer, C., Ferguson, H. C., Heckman, T. M., \& Lowenthal, J. D.1995, \apj, 454, L19

\reference{} Leitherer, C. \et 1999, \apjs, 123, 3 


\reference{} Madau, P. 1995, \apj, 441, 18

\reference{} Madau, P., Haardt, F., \& Rees, M. J. 1999, \apj, 514, 648 

\reference{} Massey, P. \& Gronwall, C. 1990, \apj, 358, 344 

\reference{} Oke, J. B., \et 1995, \pasp, 107, 375 

\reference{} Osterbrock, D. E. 1989, Physics of Gaseous Nebulae and
Active Galactic Nuclei, University Science Books, Sausalito, CA

\reference{} Ouchi, M. \et 2003, \apj, 582, 60 

\reference{} Peebles, P. J. E. 1993, Principles of Physical Cosmology, 
Princeton University Press, Princeton, New Jersey


\reference{} Rhoads, J. \et 2003, \aj, 125, 1006 

\reference{} Rhoads, J. E. \& Malhotra, S. 2001, \apj, 563, L5

\reference{} Sawicki, M., \& Yee, H. K. C. 1998, \aj, 115, 1329

\reference{} Schlegel, D. J., Finkbeiner, D. P., \& Davis, M. 1998, \apj, 500, 525 

\reference{} Shapley, A. E. \et , 2001, \apj, 562, 95

\reference{} Shapley, A. E., Steidel, C. C., Pettini, M. \& Adelberger, 
K. L. 2003, \apj, 588, 65 

\reference{} Shimasaku, K. \et 2003, \apj, 586, L111  

\reference{} Spergel, D.  \et 2003, astro-ph/0303309

\reference{} Steidel, C. C. \et 2000, \apj, 532, 170 

\reference{} Steidel, C. C., Pettini, M., \& Adelberger, K. L.  2001, \apj, 546, 665 

\reference{} Stockton, A. 1999, Astrophysics and Space Science, 269-270,
209-216

\reference{} Taniguchi, Y. \et 2003, \apj, 585, L97 

\reference{} Tenorio-Tagle, G. \et 1999, \mn, 309, 332

\reference{} Wyithe, J. S. B. \& Loeb, A. 2003, \apj, 588, L69

\reference{} Yan, H. \et 2003, \apj, 565, L93

\reference{} Yan, H. \et 2002, \apj, 580, 725

\end{references}
\end{document}